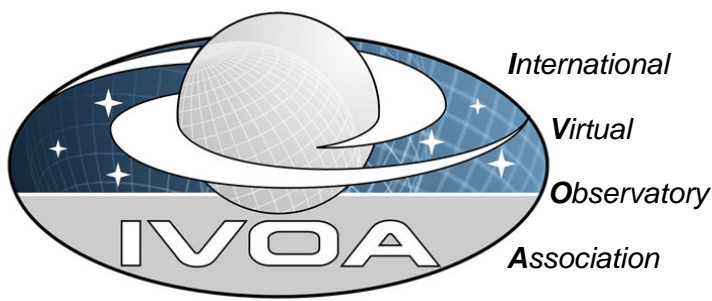

*I*nternational

*V*irtual

*O*bservatory

*A*ssociation

# Universal Worker Service Pattern Version 1.0

## IVOA Recommendation 2010-10-10

Author(s):
   P. Harrison, G. Rixon

## Abstract

The Universal Worker Service (UWS) pattern defines how to manage asynchronous execution of jobs on a service. Any application of the pattern defines a family of related services with a common service contract. Possible uses of the pattern are also described.

## Status of This Document

This document is produced by the GWS Working Group of the IVOA. The document has been reviewed by IVOA Members and other interested parties, and has been endorsed by the IVOA Executive Committee as an IVOA Recommendation as of 04 October 2010. It is a stable document and may be used as reference material or cited as a normative reference from another document. IVOA's role in making the Recommendation is to draw attention to the specification and to promote its widespread deployment. This enhances the functionality and interoperability inside the Astronomical Community.

*A list of  current IVOA Recommendations and other technical documents  can be found at http:// www.ivoa.net/Documents/.*

The words "MUST", "SHALL", "SHOULD", "MAY", "RECOMMENDED", and "OPTIONAL" (in upper or lower case) used in this document are to be interpreted as described in IETF standard, RFC 2119 [*std:rfc2119*] .

# Acknowledgements

The need for the UWS pattern was inspired by AstroGrid's Common Execution Architecture and particularly by discussions with Noel Winstanley. The ideas about statefulness are distilled from debates in the Global Grid Forum in respect of the Open Grid Services Infrastructure that was the fore-runner of Web Services Resource Framework. The REST binding came initially from suggestions by Norman Gray.

# Contents

# 1. Introduction (informative)

The Universal Worker Service (UWS) pattern defines how to build *asynchronous*, *stateful*, *job-oriented* services (the italicised terms are defined in sub-sections of this introduction). It does so in a way that allows for wide-scale reuse of software and support from software toolkits.

Section  2.  of this document describes the pattern and lists the aspects that are common to all its applications. Any such application would involve a service contract that embodies the pattern and fixes the issues left undefined in the pattern itself (see section  1.3.  ). The contract might include the XML schemata which form a "job description language" or JDL for the application, or simply a description of the input parameters for the service which are conveyed via typical HTML form encodings. It is intended that each such contract cover a family of related applications, such that the implementations may be widely reused.

Section 4 outlines several possible applications of the pattern. These use-cases may be expanded into full IVOA standards that are siblings of the current document.

Section 5 describes the preferred method of creating a synchronous facade to a UWS system.

## 1.1.  Synchronous, stateless services

Simple web services are *synchronous* and *stateless*. Synchronous means that the client waits for each request to be fulfilled; if the client disconnects from the service then the activity is abandoned. *Stateless* means that the service does not remember results of a previous activity (or, at least, the client cannot ask the service about them).

Synchronous, stateless services work well when two criteria apply.

1. The length of each activity is less than the "attention span" of the connection.
2. The results of each activity are compact enough to be easily passed back to the client via the connection on which the request was made (and possibly pushed back to the service as parameters of the next activity).

There are various limits to the attention span.

- HTTP assumes that the start of a reply quickly follows its request, even if the body of the reply takes a long time to stream. If the service takes too long to compute the results and to start the reply, then HTTP times out and the request is lost.
- A client runs on a computer which will not stay on-line indefinitely.
- A network with finite reliability will eventually break communications during an activity.
- A service is sometimes shut down for maintenance.

Synchronous, stateless services, in short, do not scale well. The following examples illustrate situations in the VO in which synchronous, stateless services are inadequate.

1. An ADQL *[std:adql]* service gives access to a large object-catalogue. Most queries run in less than a minute, but some legitimate queries involve a full-table traverse and take hours or days. The service needs to run these special cases in a low-priority queue.
2. An object-finding service runs the SExtractor *[sextractor]* application on a list of images. Normally, the list is short and the request is quickly satisfied. Occasionally, a list of 10,000 images is sent in the expectation that the work will be finished over the weekend.
3. A cone-search *[std:scs]* request on a rich catalogue raises 10,000,000 rows of results, but the client is connected via a slow link and cannot read all the results in a reasonable time. The client needs the service to send the results into storage over a faster link. This could mean sending them to VOSpace *[std:vospace]* , or simply holding them temporarily at the service until the user can retrieve them on a fast link. UWS provides a pattern for doing the latter.
4. An ADQL service allows users to save query results into new tables such that they can be the target of later queries. However, space is limited and the results tables can only be kept for a short time. The client and service negotiate the lifetime of the results tables.
5. A service performs image stacking on a list of fields. Each field can be processed by a synchronous service but the list is long and the user wants to retrieve the results of the early fields before the last fields are processed.

## 1.2.  Asynchronous and stateful services

Services can be made to scale better by making them *asynchronous* and *stateful*. Asynchronous means that a client makes two or more separate requests to the service in the course of one activity, and that the client and service may be disconnected, possibly for days or more, in between those requests. Stateful means that the service stores state information about the activity and the client addresses requests to this state.

Web services that are asynchronous are almost always stateful. Most of the special extra arrangements for asynchronous activities are actually managing the state of the activity.

There is an important class of stateful services where the state is peculiar to one job or session and the job is "owned" by one user. These, for the purpose of this document, are called *job-oriented* services. There are stateful services that are not job-oriented (e.g. a service managing a shared, client-writeable DB table), but UWS does not apply to these.

For the purpose of this discussion, let the term *job* refer to the work specified by the JDL instructions and the term *resource* refer to the state of the job as recorded by the service. Both have a finite duration. The *lifetime* of the resource – i.e. the time from inception until the service forgets the state – is generally finite and must be at least as long the duration of the job.

## 1.3.  Job description language, service contracts and universality

Consider the web-service operation that starts off a job. This operation must express what is to be done in the activity: it must carry parameters in some form.

The parameters may be expressed as a list. For example, a cone search service takes a list of three parameters: RA, DEC, RADIUS. Alternatively, the parameters may be arranged as an XML document (e.g. ADQL, CEA *[harrison05]* ). The rules for setting and arranging the parameters for a job are called the *Job-Description Language* (JDL).

The combination of the UWS pattern, a JDL and details of the job state visible to the client defines a service contract. Changing the JDL changes the contract. Thus, it is not meaningful to "implement UWS" in isolation; any implementation standard must specify the rest of the contract.

If the JDL is very general, a single service-contract can be reused for many kinds of service. AstroGrid's CEA *[harrison05]* exploits this: one JDL covers all services offering parameterised applications and even ADQL services. In the limit, a sufficiently-general JDL turns a specialized worker service into a universal worker service.

## 1.4. UWS in the IVOA Architecture

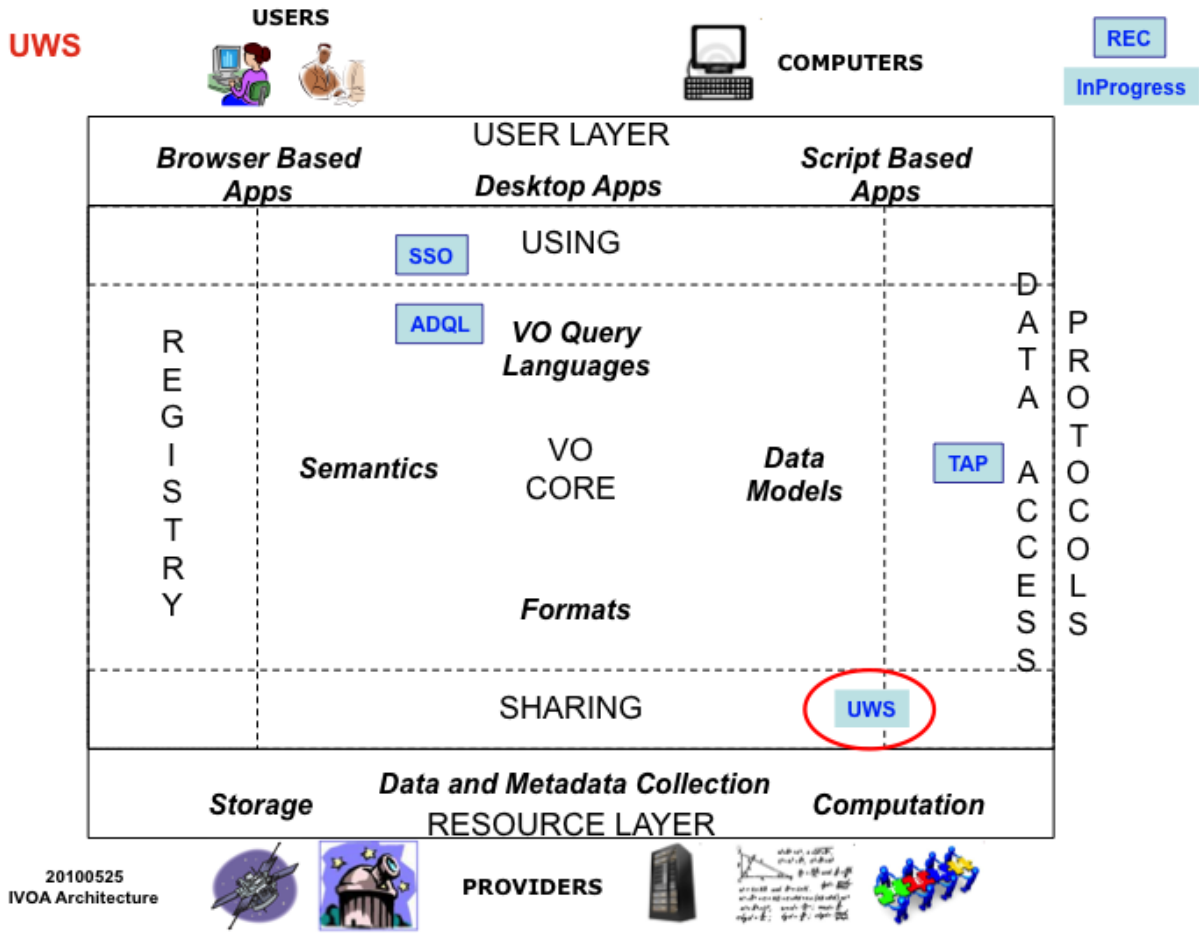

With the IVOA Architecture above, UWS is a VO infrastructure standard, being used by other standards and services to enable the development of VO applications managing asynchronous and stateless execution of jobs on VO services. Examples of such jobs include the case where the response must be computed, and that computation time takes longer than the normal expectation in an interactive web session. Additionally the UWS pattern allows a simple form of data sharing of the results of a job that is suitable for "workflow" situations and can be used by Data Access Services (currently Table Access Protocol *[std:tap]* , but potentially by other DAL standards). It utilizes IVOA standards for security *[std:ssoauth]* if it is desired that a non-public UWS be created.

# 2. Universal Worker Service Pattern (normative)

## 2.1. Objects within a UWS

A UWS consists logically of a set of objects that may be read and written to in order to control jobs. The objects are represented by elements within the XML schema detailed in Appendix B. In a REST binding, these objects are addressed as distinct web-resources each with its own URI.

The following sub-sections explain the semantics of the objects. The UML diagram below shows the relationships more succinctly.

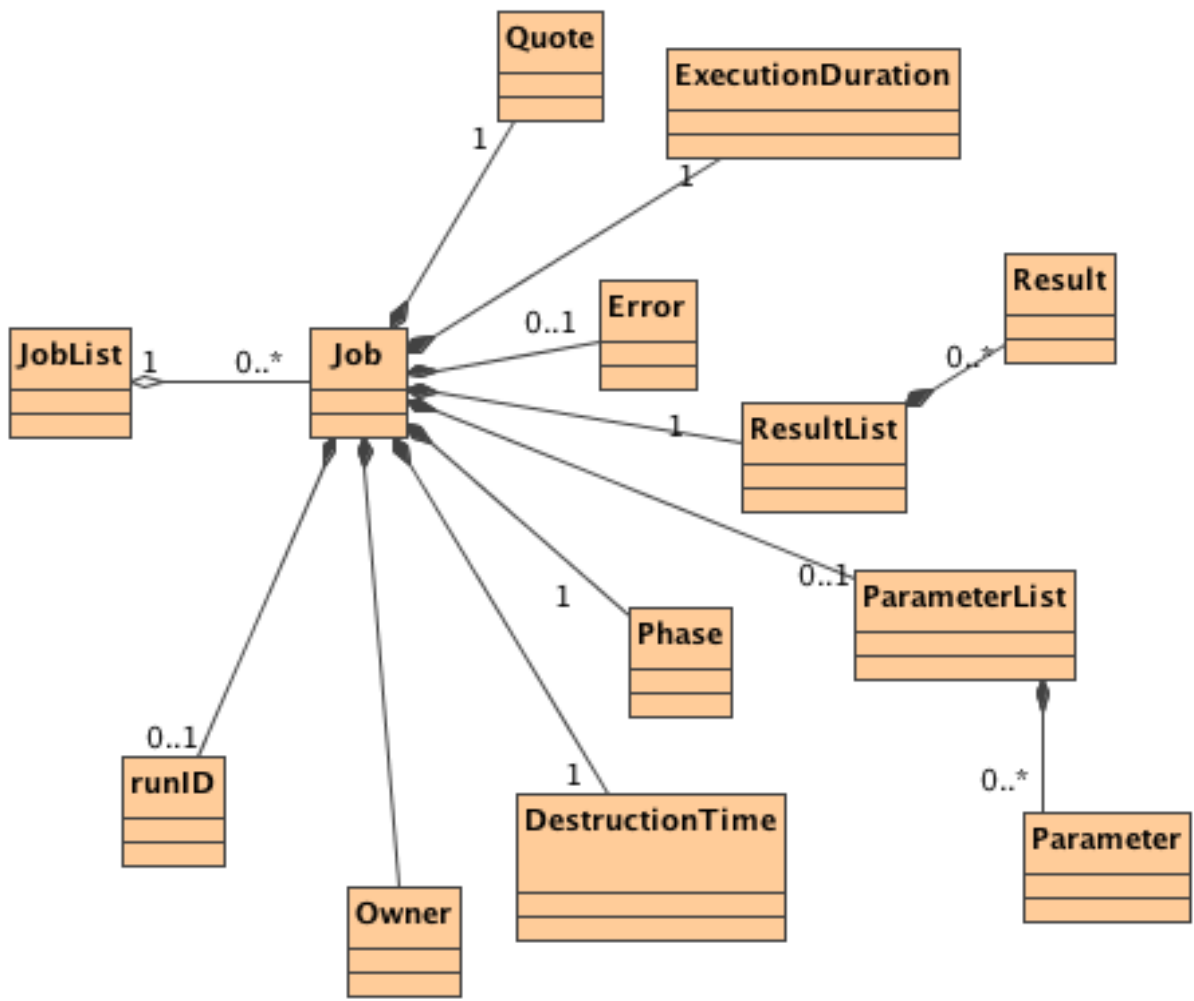

### 2.1.1. Job list

The Job List is the outermost object: it contains all the other objects in the UWS. The immediate children of the job list are Job objects (see next sub-section).

The job list may be read to find the extant jobs.

The job list may be updated to add a new job.

The job list itself does not allow jobs to be deleted. Instead, when a job is destroyed by an action on its job object, then the list updates itself accordingly.

### 2.1.2. Job

A Job object contains the state of one job. The state is a collection of other objects. Each Job contains:

- Exactly one Execution Phase.
- Exactly one Execution Duration.
- Exactly one Deletion Time
- Exactly one Quote.
- Exactly one Results List.
- Exactly one Owner.
- Zero or one Run Identifier.
- Zero or one Error.

### 2.1.3. Execution Phase

The job is treated as a state machine with the Execution Phase naming the state. The phases are:

- PENDING: the job is accepted by the service but not yet committed for execution by the client. In this state, the job quote can be read and evaluated. This is the state into which a job enters when it is first created.
- QUEUED: the job is committed for execution by the client but the service has not yet assigned it to a processor. No Results are produced in this phase.
- EXECUTING: the job has been assigned to a processor. Results may be produced at any time during this phase.
- COMPLETED: the execution of the job is over. The Results may be collected.
- ERROR: the job failed to complete. No further work will be done nor Results produced. Results may be unavailable or available but invalid; either way the Results should not be trusted.
- ABORTED: the job has been manually aborted by the user, or the system has aborted the job due to lack of or overuse of resources.
- UNKNOWN: The job is in an unknown state.
- HELD: The job is HELD pending execution and will not automatically be executed (cf, PENDING)
- SUSPENDED: The job has been suspended by the system during execution. This might be because of temporary lack of resource. The UWS will automatically resume the job into the EXECUTING phase without any intervention when resource becomes available.

A successful job will normally progress through the PENDING, QUEUED, EXECUTING, COMPLETED phases in that order. At any time before the COMPLETED phase a job may either be ABORTED or may suffer an ERROR. If the UWS reports an UNKNOWN phase, then all the client can do is re-query the phase until a known phase is reported. A UWS may place a job in a HELD phase on receipt of a PHASE=RUN request if for some reason the job cannot be immediately queued - in this case it is the responsibility of the client to request PHASE=RUN again at some later time.

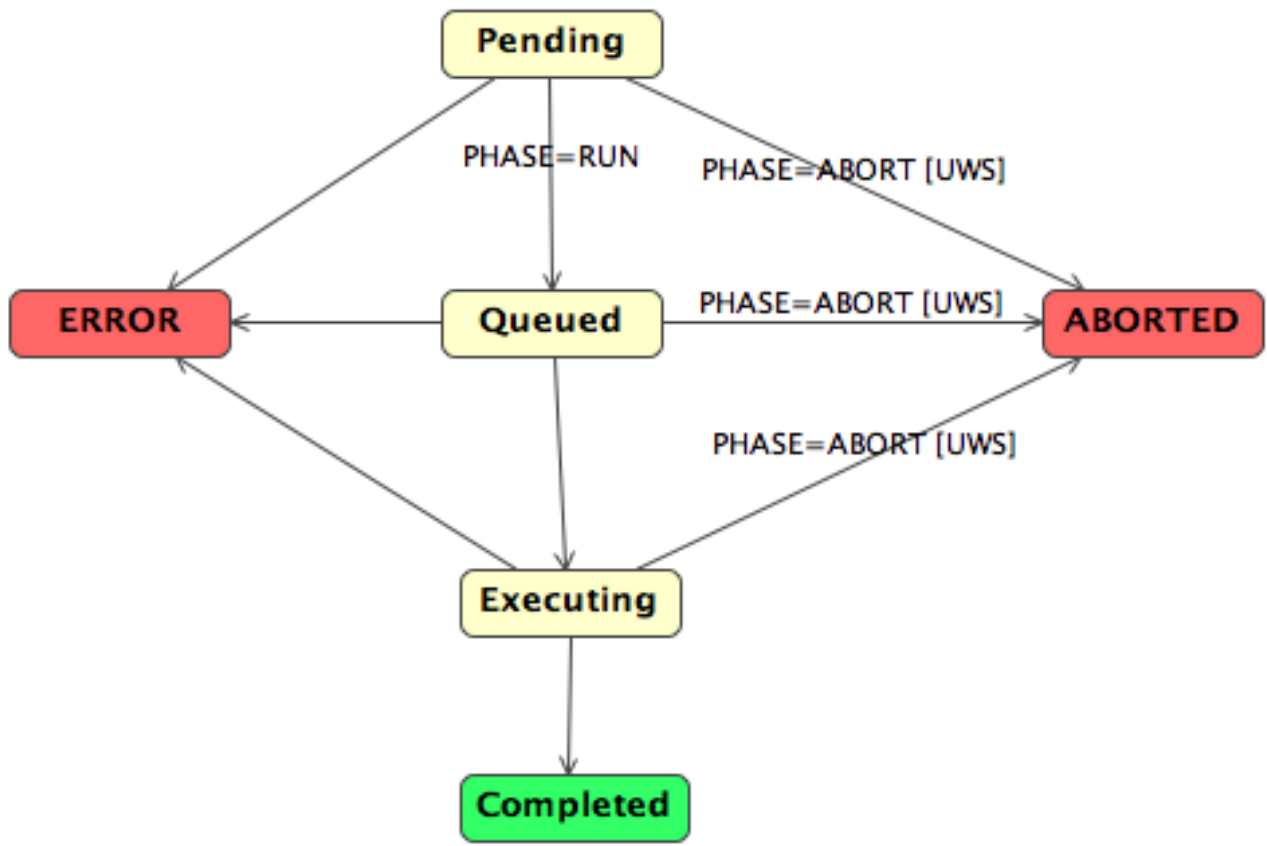

## 2.1.4. Execution Duration

An Execution Duration object defines the duration for which a job shall run. This represents the "computation time" that a job is to be allowed, although because a specific measure of CPU time may not be available in all environments, this duration is defined in real clock seconds. An execution duration of 0 implies unlimited execution duration.

When the execution duration has been exceeded the service should automatically abort the job, which has the same effect as when a manual "Abort" is requested.

Specifically, when a job is aborted:

• if the job is still executing, the execution is aborted.
• any previously generated results of the job are retained.

When a job is created, the service sets the initial execution duration. The client may write to an Execution Duration to try to change the job's cpu time allocation. The service may forbid changes, or may set limits on the allowed execution duration.

## 2.1.5. Destruction Time

The Destruction Time object represents the instant when the job shall be destroyed. The Destruction Time is an absolute time.

Destroying a job implies:

• if the job is still executing, the execution is aborted.

- any results from the job are destroyed and storage reclaimed.
- the service forgets that the job existed.

The Destruction Time should be viewed as a measure of the amount of time that a service is prepared to allocate storage for a job – typically this will be a longer duration that the amount of CPU time that a service would allocate.

When a job is created the service sets the initial Destruction Time. The client may write to the Destruction Time to try to change the life expectancy of the job. The service may forbid changes, or may set limits on the allowed destruction time.

### 2.1.6. Quote

A Quote object predicts when the job is likely to complete. The intention is that a client creates the same job on several services, compares the quotes and then accepts the best quote. From the server perspective it is possible to indicate that a job will not be run by specifying a Quote that is greater than the Destruction time.

Quoting for a computational job is notoriously difficult. A UWS implementation must always provide a quote object, in order that the two-phase committal of jobs be uniform across all UWS, but it may supply a "don't know" answer for the completion time, indicated by a negative value or an XML nil element in an XML representation of the quote object.

### 2.1.7. Error

The error object gives a human-readable error message for the underlying job. This object is intended to be a detailed error message, and consequently might be a large piece of text such as a stack trace. When there is an error running a job, a summary of the error should also be given using the optional errorSummary element of the Job element.

### 2.1.8. Owner

The owner object represents the identifier for the creator of the job. This object will not exist for all invocations of a UWS conformant protocol, but only in cases where the access to the service is authenticated as discussed more thoroughly in section 3. .

### 2.1.9. RunId

The RunId object represents an identifier that the job creator uses to identify the job. Note that this is distinct from the Job Identifier that the UWS system itself assigns to each job. The UWS system should do no parsing or processing of the RunId, but merely pass back the value (if it exists) as it was passed to the UWS at job creation time. In particular it may be the case that multiple jobs have the same RunId, as this is a mechanism by which the calling process can identify jobs that belong to a particular group. The exact mechanism of setting the RunId is not specified here, but will be part of the specification of the protocol using the UWS pattern.

### 2.1.10. Results List

The Results List object is a container for formal results of the job. Its children may be any objects resulting from the computation that may be fetched from the service when the job has completed.

Reading the Results List itself enumerates the available or expected result objects.

The children of the Results List may be read but not updated or deleted. The client cannot add anything to the Results List.

### 2.1.11. Parameter List

The Parameter List is an enumeration of the Job parameters. The form that the parameters take will depend on the JDL of the implementing service - for instance if the JDL is expressed as a single XML document, then there would be a single "parameter" that was that XML document. For services where the JDL consists of a list of name/value pairs (typical of the standard IVOA "simple" access protocols), then these would naturally be expressed in the parameter list. Each parameter value may be expressed either directly as the content of the parameter element, or the value expressed "by reference", where there returned parameter value is a URL that points to the location where the actual parameter value is stored.

A particular implementation of UWS may choose to allow the parameters to be updated after the initial job creation step, before the Phase is set to the executing state. It is up to the individual implementation to specify exactly how these parameters may be updated, but good practice (in the REST binding) would be to choose one of the following options.

1. HTTP POST an *application/x-www-form-urlencoded* parameter name, value pair to either
   a. /{jobs}/{job-id}
   b. /{jobs}/{job-id}/parameters
2. HTTP PUT the parameter value to /{jobs}/{job-id}/parameters/(parameter-name)

Additionally a particular implementation of UWS may allow the "job control" parameters (see section 2.2.3. ) to be specified as part of the JDL. If doing this then the implementation must document the possibility and must use the standard parameter names (which may cause a conflict with parameters in the JDL, which is one of the reasons for the two stage job creation pattern of UWS).

## 2.2. The REST Binding

In order to create a usable service the objects discussed in the section above must be exposed in a particular interface which can be addressed over a particular transport mechanism - this combination is known as a "binding". In this first version of the UWS pattern only a REST (Representational State Transfer) style binding  *[fielding00]* is presented, however, future versions of this document will add other bindings such as SOAP.  *[std:soap]* . It should be noted that REST is based on HTTP *[std:http]* and as such the REST binding inherits standard HTTP behaviours. In particular it should be noted that the REST binding makes use of HTTP status codes to control the behaviour of the client, especially the "300" class redirection codes to ensure that the client requests particular resources after state changing operations. If for some reason there is a failure in the underlying UWS machinery then a 500 "internal server error" status should be returned. However individual job failures are indicated by setting the appropriate parts of the job representation to error statuses and a request for the individual job object representation at /(jobs)/(jobid) should have a normal 200 status code response.

### 2.2.1. Resources and URIs

In a REST (Representational State Transfer) binding of UWS, each of the objects defined above is available as a web resource with its own URI. These URIs must form a hierarchy as shown in the table below:

| URI | Description | Representation |
|---|---|---|
| /{jobs} | the Job List | the <jobs> element in the UWS schema |
| /{jobs}/(job-id) | a Job | the <job> element in the UWS schema |
| /{jobs}/(job-id)/phase | the Phase of job (job-id) | one of the fixed strings |
| /{jobs}/(job-id)/executionduration | the maximum execution duration of (job-id) | Integer number of seconds |
| /{jobs}/(job-id)/destruction | the destruction instant for (job-id) | [*std:iso8601*] |
| /{jobs}/(job-id)/error | any error message associated with (job-id) | any representation appropriate to the implementing service |
| /{jobs}/(job-id)/quote | the Quote for (job-id) | Integer number of seconds |
| /{jobs}/(job-id)/results | any parameters for the job (job-id) | the <results> element in the UWS schema |
| /{jobs}/(job-id)/parameters | any parameters for the job (job-id) | the <parameters> element in the UWS schema |
| /{jobs}/(job-id)/owner | the owner of the job (job-id) | an appropriate identifier as discussed in  3. |

The service implementer is free to choose the names given in parentheses above, i.e. the root of the URI tree, and the form that the job identifier takes (though note that it should be a legal URI path element) - the other names are part of the UWS standard.

The URI for the Job List, in its absolute form is the root URI for the whole UWS. This URI should be given as the access URL in the UWS registration.

### 2.2.2. Representations of resources

Each of the UWS objects is mapped to a resource URI as detailed in the table above, and for each URI, a HTTP GET fetches a representation of that resource. In general where an object is a container for other objects (Job List, Job, Result List, Parameter List) then an XML representation of the object should be returned, otherwise for simple atomic types (Quote, Execution Duration etc.) a textual (mime type "text/plain") representation should be returned. The XML schema for the various objects described above is detailed in  Appendix B.   of this specification. This schema is the definitive description of the exact form of the XML returned by a UWS and must not vary between implementations.

HTTP allows multiple representations of a resource distinguished by their MIME types and selected by the HTTP "Accept" headers of a GET request. A UWS implementation can exploit this to support both web browsers and rich clients in the same tree of resources. Although the default behaviour is to return XML, a UWS could return HTML or XHTML to clients that accept these types. These clients are assumed to be web browsers and the UWS is generating its own user interface. The HTML interface generated should allow full control of the UWS via the use of HTML forms and appropriate links.

Clients which are assumed to be part of remote applications that drive UWS without showing the details to their users should accept only "application/xml,text/plain". A UWS must therefore return XML representations of the resources in preference to the HTML representation. A technique that may be used to always return XML that modern browsers can transform on the client-side to HTML is via the <?xml-stylesheet ?> processing instruction, which can be used to point to a suitable XSL resource to perform the transformation.

More detail for some of the UWS objects is provided below.

### 2.2.2.1. Job List

The representation of the Job List is a list of links to extant jobs. The list may be empty if the UWS is idle.

### 2.2.2.2. Job

The representation of a Job is as specified by the <job> element in the UWS schema as detailed in Appendix B.   an example or such a job instance is shown below

```xml
< uws:job xsi:schemaLocation = "http://www.ivoa.net/xml/UWS/v1.0 UWS.xsd " xmlns:xml = "http://www.w3.org/
XML/1998/namespace" xmlns:uws = "http://www.ivoa.net/xml/UWS/v1.0" xmlns:xlink = "http://www.w3.org/1999/
xlink" xmlns:xsi = "http://www.w3.org/2001/XMLSchema-instance" >
    < uws:jobId > cea-agdevel.jb.man.ac.uk-130.88.24.18-1242749568029-508182314 </ uws:jobId >
    < uws:ownerId xsi:nil = "true" />
    < uws:phase > COMPLETED </ uws:phase >
    < uws:startTime > 2009-05-19T17:12:48.038+01:00 </ uws:startTime >
    < uws:endTime > 2009-05-19T17:12:48.041+01:00 </ uws:endTime >
    < uws:executionDuration > 86400 </ uws:executionDuration >
    < uws:destruction > 2009-05-29T17:12:48.035+01:00 </ uws:destruction >
    < uws:parameters >
        < uws:parameter id = "scaleFactor" > 1.8 </ uws:parameter >
        < uws:parameter id = "image" byReference = "true" >  http://myserver.org/uws/jobs/jobid123/param/
image </ uws:parameter >
    </ uws:parameters >
    < uws:results >
        < uws:result id = "correctedImage" xlink:href = "http://myserver.org/uws/jobs/jobid123/result/image" />
    </ uws:results >
    < uws:errorSummary type = "transient" hasDetail = "true" >
        < uws:message > we have problem </ uws:message >
    </ uws:errorSummary >
    < uws:jobInfo >
        < any >
            < xml >
                < thatyouwant />
            </ xml >
        </ any >
    </ uws:jobInfo >
</ uws:job >
```

The <job> element has placeholders of all of the standard UWS objects, and in addition there is a <uws:jobinfo> element which can be used by implementations to include any extra information within the job description.

### 2.2.2.3. Results List

The representation of a Results List is a list of links to the resources representing the results. These resources may have any URI and any MIME type. A sensible default for their URIs is to make them children of /{jobs}/(job-id)/results, but this is not required. It may sometimes be easier for a service implementer to point to a resource on some web server separate from that running the UWS. Therefore, a client must always parse the Results List to find the results. Each result in a result list must be given a unique identifier. Where a protocol applying UWS specifies standard results it must do so fixing the identifier for those results and fixing the result URIs, however the UWS must still return a valid Results List at /{jobs}/(job-id)/results, even though in this case the identifiers and URIs could be precomputed by the client.

### 2.2.2.4. Parameters List

The representation of the parameters list is a list of elements. Each of these elements can either represent the value of the parameter directly, where the content of the element is a textual representation of the parameter, or in the case where the parameter value cannot be represented legally within XML (e.g. the parameter is a binary type such as a FITS file) then the content of the parameter is a URL to the parameter value - to indicate this case the attribute byReference is set to "true".

### 2.2.2.5. Error

When an error occurs in a job the UWS must signal this at a minimum by setting the PHASE to error. In addition the <errorSummary> element, giving a brief summary of the error, should be included within the <job> element. The UWS may include a more detailed error message for example an execution log or stack trace by providing such a resource at the /{jobs}/(jobid)/error URI. It is the responsibility of the implementing service to specify the form that such an error message may take.

### 2.2.3. State changing requests

Certain of the UWS' resources accept HTTP POST and DELETE messages to change the state of the service – This is the fundamental way that a client controls the execution of a job. In most of the cases where a job sub-object is set the response will have a http 303 "See other" status and a Location header that points back to the main job summary obtained at the /{jobs}/(job-id) URI. The job summary contains the values of (or links to) all the UWS objects within the returned XML (or XHTML). This mode of operation was chosen (as opposed to returning only the sub-object being altered) as it makes for a more natural user interface – especially in the case of the XHMTL interface. A client that wants to obtain only the value of a particular sub-object can at any time request that sub-object with a HTTP GET.

### 2.2.3.1. Creating a Job

POSTing a request to the Job List creates a new job (unless the service rejects the request). The response when a job is accepted must have code 303 "See other" and the Location header of the response must point to the created job.

This initial POST will in most cases carry parameters for the protocol that is using the UWS pattern, as detailed in  4.  .In addition for the initial post may contain job control parameters if allowed by the implementing protocol (i.e. if UWS job control parameter names do not clash with the implementing protocol parameters). One use of this facility might be to have the job placed into a potentially running state by adding PHASE=RUN to the job creation step.

### 2.2.3.2. Deleting a Job

Sending a HTTP DELETE to a Job resource destroys that job, with the meaning noted in the definition of the Job object, above. No other resource of the UWS may be directly deleted by the client. The response to this request must have code 303 "See other" and the Location header of the response must point to the Job List at the /{jobs} URI.

Posting a request with a parameter ACTION=DELETE to the Job also destroys the job, the response being as for a deletion. This action supports web browsers which typically cannot send DELETE requests.

### 2.2.3.3. Changing the Destruction Time

The Destruction Time may be changed by POSTing to /{jobs}/(job-id)/destruction. In this case, the body of the posted request is of type *application/x-www-form-urlencoded* and contains the parameter named DESTRUCTION whose value is the new destruction time in ISO8601 format; i.e. this request is what an HTML form sends.

The response to this request must have code 303 "See other" and the Location header of the response must point to the /{jobs}/(job-id) URI so that the client receives the value that the service has actually set the Destruction Time to within the Job summary response. The service may override the requested Destruction Time and substitute a value of its choosing e.g. a job owner may have reached his storage quota and so the service only allows further data to be stored for shorter times than requested.

### 2.2.3.4. Changing the Execution Duration

The Execution Duration may be changed by POSTing to /{jobs}/(job-id)/executionduration. In this case, the body of the posted request is of type *application/x-www-form-urlencoded* and contains the parameter named EXECUTIONDURATION whose value is the new executionduration in seconds.

The response to this request must have code 303 "See other" and the Location header of the response must point to the /{jobs}/(job-id) so that the client receives the value that the service has actually set the Execution Duration to. The service may to override the request and substitute a value of its choosing

### 2.2.3.5. Starting a Job

A job may be started by POSTing to the /{jobs}/(job-id)/phase URI. The POST contains a single parameter PHASE=RUN which instructs the UWS to attempt to start the job. The response to this request must have code 303 "See other" and the Location header of the response must point to the /{jobs}/(job-id) URI so that the client receives the phase that the job has been set to. Typically a UWS will put a job into a QUEUED state on receipt of this command, but depending on how busy the server is, the job might be put almost immediately (and without client intervention) into an EXECUTING state.

### 2.2.3.6. Aborting a Job

A job may be aborted by POSTing to the /{jobs}/(job-id)/phase URI. The POST contains a single parameter PHASE=ABORT which instructs the UWS to attempt to abort the job. Aborting a job has the effect of stopping a job executing, but the resources associated with a job remain intact. The response to this request must have code 303 "See other" and the Location header of the response must point to the /{jobs}/(job-id) URI so that the client receives the phase that the job has been set to.

### 2.2.4. Message pattern

The REST binding results in the message pattern shown in figure 2.

*Illustration 2: Typical calling sequence for the REST binding of UWS*

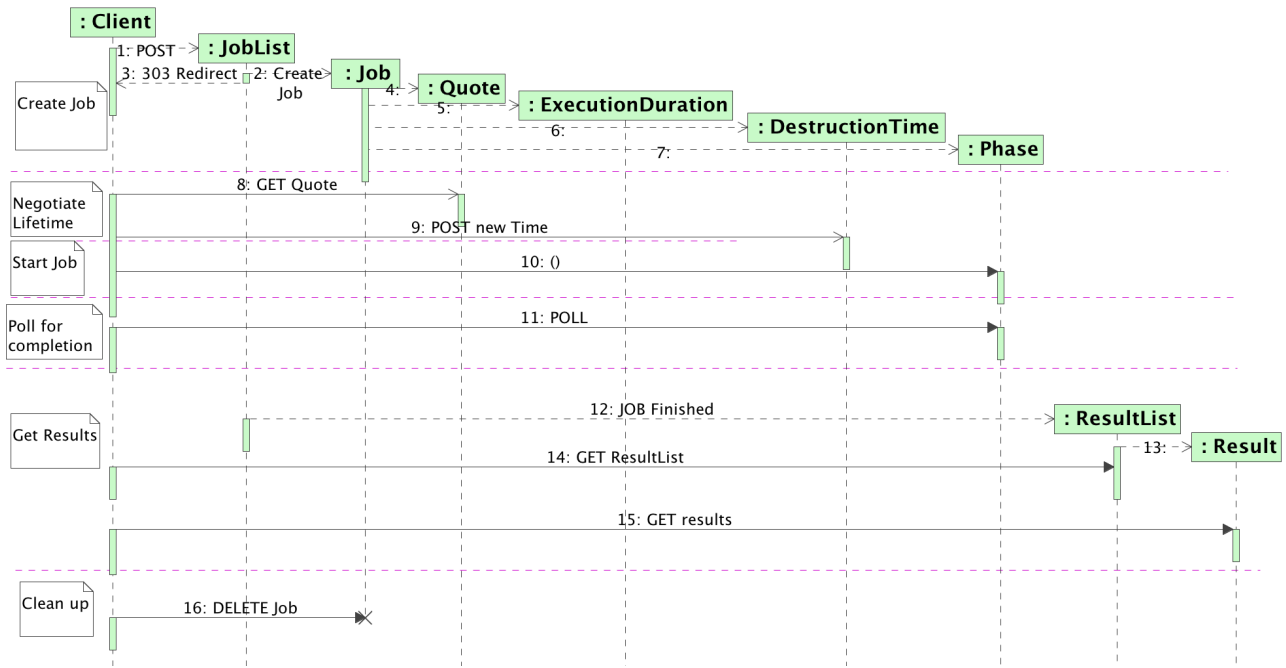

## 3. Security Considerations (normative)

A UWS should follow IVOA standards for security *[std:ssoauth]* if it is desired that a non-public UWS be created. It is possible to imagine many different authorization policies that might be employed in an authenticated UWS, where particular groups of users have different permissions to create and view different types of jobs. A full discussion of such authorization policies is beyond the scope of this document, but a UWS should behave as described in section 2.2. for any individual authenticated user, although it is clear that a user without sufficient privileges might only obtain a restricted list of jobs within the joblist at /{jobs}. Any attempt to retrieve a job for which the user does not have sufficient privilege should result in a 403 Forbidden HTTP status being returned.

When an authentication mechanism is used in the UWS then the implementation should set the owner object to the identity obtained by the authentication.

## 4. Applications of UWS (informative)

The UWS pattern leaves undefined two essential parts of the service contract: the content that must be posted to create a job; and the pattern of results made available by a completed job. An application of UWS completes a service contract by defining these matters.

There follow some use cases applying the UWS pattern. The descriptions are neither formal nor complete. The intention is to show a range of ways that the pattern can be applied without burdening the reader with the level of detail needed for a standard implementation.

Any of these cases could be worked up into a full IVOA standard by formalizing the description, adding detail and generally making the specification more exact and complete.

## 4.1. Image service with data staging

Consider a service that computes images from archive data. The computation takes significant time, so the service is asynchronous. The service keeps the computed images in its own storage until the user downloads them; this is essentially the model of SIAP *[std:siap]* .

The asynchronous image-service is a logical extension of a synchronous SIA service. Therefore it uses the REST binding of UWS.

The parameters for posting a new job are as for SIAP 1.0:

- POS, the position on the sky to be searched.
- SIZE, the size of the search box.
- FORMAT: the type of images to be computed.

Particular implementations are free to add extra parameters.

These parameters are posted in a document of type *application/x-www-form-urlencoded*: i.e. they can be sent from an HTML form.

The images generated by the job are accessible as results. Each image has its own URI and can be downloaded over HTTP at any time until the destruction time of the job. The URIs for the images may be discovered from the Results List in the normal UWS way.

SIAP 1.0 produces, for each query, a table of metadata describing the images. The asynchronous image-service produces a table to the same schema as a result with a fixed identifier, called "table".

Image results are added to the results list, and to the "table" result, as they are generated. Hence, a client that polls the service can discover, download and use some of the images before the job is finished. If the client is satisfied with these early images, the client can cancel the rest of the job by destroying the job. However, destroying the job deletes the cached images so the client has to download them *first*.

## 4.2. ADQL service with cursor

ADQL [1] can serve as a JDL. Consider an ADQL service that supports long-running queries as asynchronous operations. In general, the results of the query may be a large set of data. They may be too large to download comfortably. We might like to cache these results on the service and to operate a cursor, drawing down from the resource a few rows of the table at a time.

The parameters of a job are as follows:

- ADQL: the query text
- FORMAT: the format for the results

These parameters are posted in a document of type *application/x-www-form-urlencoded*: i.e. they can be sent from an HTML form.

A successful query generates the following results:

- *table*: the whole result set as one file resource.
- *header*: the metadata for the output table.
- *cursor*: a selection of rows of output.

The *cursor* result is parameterised by the query parameters FIRST and LAST in the query string of its URI: these parameters state the index of the first and last row to be returned; e.g.

http://whatever.org/adlqService/results/cursor?FIRST=1&LAST=100

If the parameters are missing, the service decides which rows to return.

## 4.3. Parameterised applications

There is a class of applications on which a job may be defined by a list of simple parameters. "Simple" here means unstructured: a scalar value such as a number, a string of text or a boolean value. If the parameters are allowed to be file name, so that structured data are passed indirectly, then the class of applications is very large indeed. Almost any non-interactive application can be driven in this way.

Turning each application of choice into a service (with or without UWS semantics) would be onerous. However, if the application's interface is entirely characterized, through the JDL, in terms of typed input and output parameters, then one service contract will work for all the applications and a single implementation of the contract can be reused for all cases.

AstroGrid's Common Execution Architecture (CEA) *[harrison05]* works in this way. It has just one service contract for all applications (including ADQL services; the ADQL query is passed in the list of parameters). It has four implementations, one for each of the possible interfaces between the service and a kind of job (jobs can be implemented with Java classes, command-line applications, HTTP-get services or JDBC databases). CEA also specifies stateful, asynchronous services and makes use of VOSpace.

Consider a CEA reworked to use the UWS pattern for consistency with other (future) IVOA standards. Call it CEA v2 to distinguish it from CEA v1 as currently maintained by AstroGrid. For this example, consider the particular kind of CEA service that runs applications supplied as executable binaries.

A binary application-server has a library of applications co-located with its service and defined in the service configuration set by the service provider. It does not accept code from the client for local execution.

The JDL in CEA v2 is similar to that in CEA v1 It is a formal, XML vocabulary for expressing choice of application and parameter lists *[std:vocea]* . Parameters may be inputs or outputs of the job.

To start a job, a document in this JDL is posted to the UWS. The document is sent in its native MIME-type, application/xml, so this is not an interface that can be driven directly from an HTML form, although it can be driven from the emerging XForms technology *[std:xform]* .

The results of the job depend on the choice of application. They are all results whose identifiers and types are defined in the definition of the application. That application-definition is registered, so the client knows before running the job what results to expect.

CEA input-parameters may be indirect: i.e. they may refer to data in on-line storage. In this case, the JDL document contains the URIs for those data objects Alternatively, the parameters may be direct, in which case the JDL contains the actual value of the parameters.

Similarly, CEA results may be made indirect. In this case, the results are named as parameters in the JDL where the values are the URIs to which the results are delivered. The application server can then stream the results to the specified destination as they become available and need not cache them locally. If a job result is indirect, then the server can choose whether or not to keep a local copy. If it chooses not to cache locally, then the result URI in the UWS is set to the external location named in the URI.

# 5. Implementing a Synchronous Service on top of UWS (informative)

Whilst the provision of synchronous services is not one of the design aims of the UWS pattern, there is clearly a desire in second generation IVOA services still to allow a simple synchronous calling pattern to be available to simple clients. What follows is a recommended recipe for putting a synchronous facade on UWS;

It is assumed that the core of the service does provide a true UWS compliant set of endpoints rooted at /async (equivalent to the /{jobs} endpoint in the nomenclature used above). The desired synchronous service is to be rooted at /sync.

1. The job is started by either a GET (for compatibility with existing IVOA standards) or a POST (preferably) of form encoded parameters to the /sync endpoint.
2. Internally the service creates a job in the standard UWS system with the given parameters and sets the PHASE to RUN, noting the returned job identifier which we will call {job-id}. The /sync endpoint then responds with a STATUS 303 (redirection) response to the URL /sync/{job-id}.
3. The /sync/{job-id} endpoint then blocks until it detects that the underlying job has finished at which point it responds with a STATUS 303 (redirection) to the /{jobs}/{job-id}/results/mainresult URL, where "mainresult" is the name of the primary result of the job.

In this way the service appears to be a synchronous to the original client – assuming it obeys standard HTTP redirection semantics, so a simple client like a web browser could obtain the result with a single "click". At the same time a more sophisticated, UWS aware, client could control the same job from the standard /{jobs} endpoint – indeed if the synchronous call timed out for some reason, then it would be possible for the original client to retrieve the results by looking at the /{jobs}/{job-id} URL tree, because it could make the association of the job-Id from the URL it receives in step 2 above.

The purely synchronous client is restricted compared with the full UWS pattern in that there can only be one result directly returned to the client, as noted in stage 3 above. This is usually not a problem for compatibility with existing version 1.0 DAL services as they typically return a single VOTable containing references to the desired data.

# Appendices

# Appendix A.  Updates from previous versions

## Appendix A.1. At Version 0.5

- changed the POST parameter names to be the same as the resource paths.
- added synchronous section.
- updated SOAP binding section

## Appendix A.2. At Version 1.0

- Removed all SOAP binding - deferred to later version.
- Added parameterList
- Added ownerId and jobId as subsidiary job object

# Appendix B. UWS Schema

Note that this schema can be found on-line at http://www.ivoa.net/xml/UWS/v1.0 (i.e. the target namespace can also be used as a URL for the schema.) This location should represent the definitive source, the schema is reproduced below only for completeness of this document.

```xml
<!-- $Revision: 1353 $ $Date: 2010-10-05 15:09:40 +0100 (Tue, 05 Oct 2010) $ $HeadURL:
https://volute.googlecode.com/svn/trunk/projects/grid/uws/doc/UWS.html $ -->
<!-- UWS schema - Paul Harrison May 2008 -->
<!-- $Log: UWS.xsd,v $ Revision 1.1.2.1 2010/02/25 11:33:41 pah updates as used in the UWS
1.0 PR Revision 1.1 2009/06/15 15:30:32 pah made some of the global element defintions
for job suboobjects just local again added runid, ownerid added ParamList changed some
element names to be consistent with the uris Revision 1.6 2008/10/01 11:54:02 pah fix
up cvs header Revision 1.5 2008/09/25 00:22:35 pah change termination time to execution
duration -->
<xs:schema targetNamespace = "http://www.ivoa.net/xml/UWS/v1.0" elementFormDefault = "qualified"
    attributeFormDefault = "unqualified" xmlns:xml = "http://www.w3.org/XML/1998/namespace" xmlns:xs =
    "http://www.w3.org/2001/XMLSchema" xmlns:uws = "http://www.ivoa.net/xml/UWS/v1.0" xmlns:xlink = "http://
    www.w3.org/1999/xlink" >
    <!-- <xs:import namespace="http://www.w3.org/1999/xlink" schemaLocation="../../../stc/
    STC/v1.30/XLINK.xsd" /> -->
    <xs:import namespace = "http://www.w3.org/1999/xlink" schemaLocation = "http://www.ivoa.net/xml/Xlink/
    xlink.xsd" />
    <xs:complexType name = "ShortJobDescription" >
        <xs:sequence >
            <xs:element name = "phase" type = "uws:ExecutionPhase" >
                <xs:annotation >
                    <xs:documentation >  the execution phase - returned at /(jobs)/(jobid)/phase </
                    xs:documentation >
                </xs:annotation >
            </xs:element >
        </xs:sequence >
        <xs:attribute name = "id" type = "uws:JobIdentifier" use = "required" />
        <xs:attributeGroup ref = "uws:reference" />
    </xs:complexType >
    <xs:attributeGroup name = "reference" >
        <xs:annotation >
            <xs:documentation > standard xlink references  </xs:documentation >
        </xs:annotation >
        <xs:attribute ref = "xlink:type" use = "optional" default = "simple" />
        <xs:attribute ref = "xlink:href" use = "optional" />
    </xs:attributeGroup >
    <xs:simpleType name = "ExecutionPhase" >
        <xs:annotation >
            <xs:documentation >  Enumeration of possible phases of job execution </xs:documentation >
        </xs:annotation >
        <xs:restriction base = "xs:string" >
            <xs:enumeration value = "PENDING" >
                <xs:annotation >
                    <xs:documentation >  The first phase a job is entered into - this is where a job is
                    being set up but no request to run has occurred. </xs:documentation >
                </xs:annotation >
            </xs:enumeration >
            <xs:enumeration value = "QUEUED" >
                <xs:annotation >
                    <xs:documentation >  An job has been accepted for execution but is waiting in a
                    queue </xs:documentation >
                </xs:annotation >
            </xs:enumeration >
            <xs:enumeration value = "EXECUTING" >
                <xs:annotation >
                    <xs:documentation >  An job is running </xs:documentation >
                </xs:annotation >
            </xs:enumeration >
```

```xml
                < xs:enumeration value = "COMPLETED" >
                    < xs:annotation >
                        < xs:documentation >  An job has completed successfully  </ xs:documentation >
                    </ xs:annotation >
                </ xs:enumeration >
                < xs:enumeration value = "ERROR" >
                    < xs:annotation >
                        < xs:documentation >  Some form of error has occurred  </ xs:documentation >
                    </ xs:annotation >
                </ xs:enumeration >
                < xs:enumeration value = "UNKNOWN" >
                    < xs:annotation >
                        < xs:documentation >  The job is in an unknown state.  </ xs:documentation >
                    </ xs:annotation >
                </ xs:enumeration >
                < xs:enumeration value = "HELD" >
                    < xs:annotation >
                        < xs:documentation >  The job is HELD pending execution and will not automatically
                        be executed - can occur after a PHASE=RUN request has been made (cf PENDING).
                        </ xs:documentation >
                    </ xs:annotation >
                </ xs:enumeration >
                < xs:enumeration value = "SUSPENDED" >
                    < xs:annotation >
                        < xs:documentation >  The job has been suspended by the system during execution
                        </ xs:documentation >
                    </ xs:annotation >
                </ xs:enumeration >
                < xs:enumeration value = "ABORTED" >
                    < xs:annotation >
                        < xs:documentation >  The job has been aborted, either by user request or by the
                        server because of lack or overuse of resources. </ xs:documentation >
                    </ xs:annotation >
                </ xs:enumeration >
            </ xs:restriction >
    </ xs:simpleType >
    < xs:complexType name = "JobSummary" >
        < xs:annotation >
            < xs:documentation > The complete representation of the state of a job </ xs:documentation >
        </ xs:annotation >
        < xs:sequence >
            < xs:element name = "jobId" type = "uws:JobIdentifier" />
            < xs:element name = "runId" type = "xs:string" maxOccurs = "1" minOccurs = "0" >
                < xs:annotation >
                    < xs:documentation >  this is a client supplied identifier - the UWS system
                    does nothing other than to return it as part of the description of the job </
                    xs:documentation >
                </ xs:annotation >
            </ xs:element >
            < xs:element name = "ownerId" type = "xs:string" nillable = "true" >
                < xs:annotation >
                    < xs:documentation > the owner (creator) of the job - this should be expressed as a
                    string that can be parsed in accordance with IVOA security standards. If there was no
                    authenticated job creator then this should be set to NULL. </ xs:documentation >
                </ xs:annotation >
            </ xs:element >
            < xs:element name = "phase" type = "uws:ExecutionPhase" >
                < xs:annotation >
                    < xs:documentation >  the execution phase - returned at /(jobs)/(jobid)/phase </
                    xs:documentation >
                </ xs:annotation >
            </ xs:element >
            < xs:element name = "quote" type = "xs:dateTime" nillable = "true" maxOccurs = "1"
            minOccurs = "0" >
                < xs:annotation >
```

```xml
                < xs:documentation >  A Quote predicts when the job is likely to complete - returned
          at /(jobs)/(jobid)/quote "don't know" is encoded by setting to the XML null value
          xsi:nil="true" </ xs:documentation >
      </ xs:annotation >
  </ xs:element >
  < xs:element name = "startTime" type = "xs:dateTime" nillable = "true" >
          < xs:annotation >
                < xs:documentation > The instant at which the job started execution. </
                xs:documentation >
          </ xs:annotation >
  </ xs:element >
  < xs:element name = "endTime" type = "xs:dateTime" nillable = "true" >
          < xs:annotation >
                < xs:documentation > The instant at which the job finished execution </
                xs:documentation >
          </ xs:annotation >
  </ xs:element >
  < xs:element name = "executionDuration" type = "xs:int" nillable = "false" >
  <!-- TODO look if xs:duration here has any benefits -->
          < xs:annotation >
                < xs:documentation > The duration (in seconds) for which the job should be
                allowed to run - a value of 0 is intended to mean unlimited - returned at /(jobs)/(jobid)/
                executionduration </ xs:documentation >
          </ xs:annotation >
  </ xs:element >
  < xs:element name = "destruction" type = "xs:dateTime" nillable = "true" >
          < xs:annotation >
                < xs:documentation >  The time at which the whole job + records + results will be
                destroyed. returned at /(jobs)/(jobid)/destruction </ xs:documentation >
          </ xs:annotation >
  </ xs:element >
  < xs:element ref = "uws:parameters" maxOccurs = "1" minOccurs = "0" >
          < xs:annotation >
                < xs:documentation > The parameters to the job (where appropriate) can also be
                retrieved at /(jobs)/(jobid)/parameters </ xs:documentation >
          </ xs:annotation >
  </ xs:element >
  < xs:element ref = "uws:results" >
          < xs:annotation >
                < xs:documentation > The results for the job - can also be retrieved at /(jobs)/(jobid)/
                results </ xs:documentation >
          </ xs:annotation >
  </ xs:element >
  < xs:element name = "errorSummary" type = "uws:ErrorSummary" maxOccurs = "1" minOccurs
  = "0" >  </ xs:element >
  < xs:element name = "jobInfo" maxOccurs = "1" minOccurs = "0" >
          < xs:annotation >
                < xs:documentation >  This is arbitrary information that can be added to the job
                description by the UWS implementation. </ xs:documentation >
          </ xs:annotation >
          < xs:complexType >
                < xs:sequence >
                      < xs:any namespace = "##any" processContents = "lax" minOccurs = "0"
                      maxOccurs = "unbounded" />
                </ xs:sequence >
          </ xs:complexType >
  </ xs:element >
          </ xs:sequence >
  </ xs:complexType >
  < xs:simpleType name = "JobIdentifier" >
          < xs:annotation >
                < xs:documentation >  The identifier for the job  </ xs:documentation >
          </ xs:annotation >
          < xs:restriction base = "xs:string" />
  </ xs:simpleType >
  < xs:element name = "job" type = "uws:JobSummary" >
```

```
        < xs:annotation >
            < xs:documentation >  This is the information that is returned when a GET is made for a single
                job resource - i.e. /(jobs)/(jobid) </ xs:documentation >
        </ xs:annotation >
</ xs:element >
< xs:element name = "jobs" >
        < xs:annotation >
            < xs:documentation >  The list of job references returned at /(jobs) </ xs:documentation >
        </ xs:annotation >
        < xs:complexType >
            < xs:annotation >
                < xs:documentation >  ISSUE - do we want to have any sort of paging or selection
                    mechanism in case the job list gets very large? Or is that an unnecessary complication... </
                    xs:documentation >
            </ xs:annotation >
            < xs:sequence >
                < xs:element name = "jobref" type = "uws:ShortJobDescription" maxOccurs =
                    "unbounded" minOccurs = "0" />
            </ xs:sequence >
        </ xs:complexType >
</ xs:element >
< xs:complexType name = "ResultReference" >
        < xs:annotation >
            < xs:documentation >  A reference to a UWS result  </ xs:documentation >
        </ xs:annotation >
        < xs:attribute name = "id" type = "xs:string" use = "required" />
        < xs:attributeGroup ref = "uws:reference" />
</ xs:complexType >
< xs:element name = "results" >
        < xs:annotation >
            < xs:documentation >  The element returned for /(jobs)/(jobid)/results </ xs:documentation >
        </ xs:annotation >
        < xs:complexType >
            < xs:sequence >
                < xs:element name = "result" type = "uws:ResultReference" maxOccurs = "unbounded"
                    minOccurs = "0" />
            </ xs:sequence >
        </ xs:complexType >
</ xs:element >
< xs:complexType name = "ErrorSummary" >
        < xs:annotation >
            < xs:documentation >  A short summary of an error - a fuller representation of the error may be
                retrieved from /(jobs)/(jobid)/error  </ xs:documentation >
        </ xs:annotation >
        < xs:sequence >
            < xs:element name = "message" type = "xs:string" />
        </ xs:sequence >
        < xs:attribute name = "type" type = "uws:ErrorType" use = "required" >
            < xs:annotation >
                < xs:documentation >  characterization of the type of the error  </ xs:documentation >
            </ xs:annotation >
        </ xs:attribute >
        < xs:attribute name = "hasDetail" type = "xs:boolean" use = "required" >
            < xs:annotation >
                < xs:documentation > if true then there is a more detailed error message available at /
                    (jobs)/(jobid)/error </ xs:documentation >
            </ xs:annotation >
        </ xs:attribute >
</ xs:complexType >
< xs:simpleType name = "ErrorType" >
        < xs:restriction base = "xs:string" >
            < xs:enumeration value = "transient" />
            < xs:enumeration value = "fatal" />
        </ xs:restriction >
</ xs:simpleType >
< xs:complexType name = "Parameter" mixed = "true" >
```

```
            < xs:annotation >
                < xs:documentation >  the list of input parameters to the job - if the job description language
                does not naturally have parameters, then this list should contain one element which is the content
                of the original POST that created the job.  </ xs:documentation >
        </ xs:annotation >
        < xs:attribute name = "byReference" type = "xs:boolean" default = "false" >
            < xs:annotation >
                < xs:documentation >  if this attribute is true then the content of the parameter represents
                a URL to retrieve the actual parameter value. </ xs:documentation >
                < xs:documentation >  It is up to the implementation to decide if a parameter value cannot
                be returned directly as the content - the basic rule is that the representation of the parameter
                must allow the whole job element to be valid XML. If this cannot be achieved then the
                parameter value must be returned by reference. </ xs:documentation >
            </ xs:annotation >
        </ xs:attribute >
        < xs:attribute name = "id" type = "xs:string" use = "required" >
            < xs:annotation >
                < xs:documentation >  the identifier for the parameter  </ xs:documentation >
            </ xs:annotation >
        </ xs:attribute >
        < xs:attribute name = "isPost" type = "xs:boolean" />
    </ xs:complexType >
    < xs:element name = "parameters" >
        < xs:complexType >
            < xs:sequence >
                < xs:element name = "parameter" type = "uws:Parameter" maxOccurs = "unbounded"
                minOccurs = "0" />
            </ xs:sequence >
        </ xs:complexType >
    </ xs:element >
</ xs:schema >
<!-- -->
```